\documentclass{article}
\usepackage[utf8]{inputenc}
\usepackage{graphicx}
\usepackage[backend=bibtex,style=numeric]{biblatex}
\graphicspath{ {./images/} }

\begin{document}

\title{An Exploration of Mars Colonization with Agent-Based Modeling}
\author{
Edgar Arguello \footnote{Authors are listed in alphabetical order.}
\and Sam Carter
\and Cristina Grieg
\and Michael Hammer
\and Chris Prather
\and Clark Petri 
\and Anamaria Berea\footnote{
Department of Computational Social Sciences, George Mason University}
}
\date{}

\maketitle

\section{Introduction}
Establishing a human settlement on Mars is an incredibly complex engineering problem. The inhospitable nature of the Martian environment requires any habitat to be largely self-sustaining. Beyond mining a few basic minerals and water \cite{Clery}, the colonizers will be dependent on earth resupply and replenishment of necessities via technological means, i.e., splitting martian water into oxygen for breathing and hydrogen for fuel \cite{Dunnill}.

Our goal is to better understand the interactions of future Martian colonists through an Agent-Based Modeling (ABM) approach. Accounting for engineering and technological limitations, we draw on research regarding high performing teams in isolated and high stress environments (ex: submarines, Arctic exploration, war) to include psychological components within the ABM. Interactions between agents with different psychological profiles are modeled at the individual level while global events such as accidents or delays in earth resupply affect the colony as a whole.

\section{Space Settlements Literature Review}
\subsection{Use of Agent-Based Models to Simulate Small-Scale Communities}
Given the many facets to the problem, a complex systems approach is appropriate to extend our insights and avoid uninformed predictions later \cite{Kohler}. ABMs may prove useful to help understand the ecological and demographic conditions needed for sustainable resource-management in a constrained human habitat \cite{Kohler}. It has been demonstrated that integrating Geographic Information System (GIS) overlays and multi-agent simulation provides a means to explore group behaviors which may be affected by an agent's spatial knowledge of their terrain \cite{Book}. ABMs provide useful “laboratories” for competing hypotheses to identify rules of behavior that lead to group dynamics \cite{Book}, which we would then like to apply to our simulation of a potential Mars settlement.

\subsection{Team and Interpersonal Psychology}
The modern era of teams working in extreme conditions commenced with the advent of nuclear submarines in the 1950s, the space race of the 1960s, and expanded Arctic exploration of the 1900s \cite{Driskell}. These pursuits specifically buoyed interest in how teams perform in extreme environments, what makes them successful, and how individuals cope with the challenges presented by such conditions. The technical sophistication necessary for persistent survival in submarine missions or space exploration, coupled with their isolation and inherent risks, position those endeavors as proxies for initializing a martian colony ABM.

In a study of Antarctic exploration teams, researchers found that psychological discomfort that occurred was limited to the individual or between individuals levels and not the group as a whole \cite{Wood}. Furthermore, the researchers identified a steady linear increase in group tension over the course of the mission. The interplay of individual interactions as well as a broad and continual drain on psychological well-being, albeit minor, are concepts which underpin much of our psychological modeling. It should be noted that specific personality profiles as determined by psychological examination were predictive of interpersonal cohesion in some instances \cite{Wood}, though these considerations are an area for future research.

Attempts to connect quantitative biomarkers such as cortisol to psychological examination performance while operating in extreme environments have shown inconclusive results \cite{Sandal}. However, the same study did identify overarching coping mechanisms of a submarine crew. While these mechanisms showed no correlation to the personality profiles, they were beneficial. Traits such as strong achievement motivation and interpersonal orientation were associated with superior coping during submarine missions \cite{Sandal}. In another study of submarine service, optimism, humor accompanied by cynicism, and a strong perception of the submarine service were the primary coping mechanisms exhibited by the crew \cite{Kimhi}. Additionally, having some semblance of occasional privacy with their own bed was a highly valuable coping mechanism, a consideration for future research with more granular spatial modeling.

Research conducted on Soldiers and Marines during their deployment to Iraq demonstrated that the intensity of combat situations is the primary determinant of behavioral health challenges \cite{Castro}. Other deployment-issues, such as their length and frequency, also impacted the psychological health of the studied Soldiers and Marines, albeit to a lesser extent than direct combat.

The present literature demonstrates that team and individual psychological success in extreme environments can be broadly attributed to coping capacity, which we define as the \textit{ability of people, organizations and systems, using available skills and resources, to manage adverse conditions, risk or disasters}. Within the model, coping capacity is aggregated from a series of sub-categories and quantitatively incorporated on a Cronbach alpha scale for each individual agent. Cronbach’s alpha scoring is drawn from the psychological changes in a 100-Day remote field groups study, where researchers examined two Australian teams traversing across the Lambert Glacier Basin, paying close attention to personality characteristics, environmental factors, and interpersonal factors as predictors of Group Tensions, Personal Morale, Emotional State, Cognitive Readiness, and the Team’s Work-Life \cite{Wood}. 

The agents within the model are categorized into four separate personality categories: Agreeable, Social, Reactive, and Neurotic. Their coping ability and resilience are linked to personality profiles typified by strong instrumentality and achievement motivation combined with interpersonal sensitivity. The four categories are drawn from the \textit{Positive Instrumental-Expressive}, the \textit{Hostile, Competitive Interpersonal orientation}, and the \textit{low achievement motivation combined with passive-aggressive characteristics} linked to Personality Characteristics Inventory (PCI) profiles which NASA has used for screening, monitoring, and evaluating subjects in isolated and confined extreme (ICE) environments \cite{Sandal}. The categorical measurements are as follows: 

\begin{itemize}
    \item Agreeables - Individuals with the lowest degree of competitiveness, low aggressiveness, and not fixated of stringent routine. Cronbach Alpha score starting at 0.97
    \item Socials - Individuals with a medium degree of competitiveness, extroverted, require social interaction, but are not fixated on stringent routines. Cronbach Alpha score starting at 0.94
    \item Reactives - Individuals with a medium degree of competitiveness, competitive interpersonal orientation, and fixated on stringent routines. Cronbach Alpha score starting at 0.89
    \item Neurotics - Individuals with a high degree of competitiveness, highly aggressive interpersonal characteristics, and challenged ability to adapt to boredom or a change in routine. Cronbach Alpha score starting at 0.84
\end{itemize}

Additionally, each agent is granted skills associated with their civilian and military occupational specialties consistent with NASA’s Human Factors and Behavioral Performance Element research, which analyzed the abilities that are generalizable across circumstances and crew roles and those that will be required by all crew members during a 30-month expedition to Mars \cite{Stuster}.

The skills are broken up into two categories; Management (Skill 1) and Engineer (Skill 2), which is extrapolated from NASA’s more comprehensive chart of Leader, Pilot, Physician, Biologist, Geologist, Computer, Electrician, Mechanic, and Combined \cite{Stuster}. The psychological structure of the model is explored in more detail in section 3.

\subsection{Martian Environment and Local Economy}
Our challenge is to examine the unique institutional conditions required to manage a settlement at such distance from Earth.  In particular, we sought to address the following questions: What conditions are needed to maintain a stable colony on Mars? What combination of personality types would do best in this hostile environment? How many resources are needed for two years between resupplies and assuming occasional accidents?  To tackle these questions, we examined research relating to similar questions in four contexts: existing ABMs, economic and related literature, ISS and Antarctica outpost data, and agricultural research.

Collectively, the research suggests important design principles for Martian economic modeling.  Economic modeling of a remote settlement involves modeling the external dependencies (food, etc.) and the internal labor and trade economy of the settlement itself.  Food is the single largest external dependency.  For example, food shipments arrive at the McMurtry and other Antarctica settlements only once annually, so this provides us with a starting point for calculating the estimated bi-annual shipments to Mars.  Consequently, the frequency and size of external food shipment, as well as the degree to which a Mars settlement would attempt to grow food, are essential variables to calibrate to model a sustainable Mars settlement.

Second, waste disposal and recycling (including carbon dioxide from humans and other types of emissions from plants) are costs that must be included. Our current focus is on supporting a stable population on Mars, so waste disposal, recycling, economic gain from mining minerals, and other second order considerations were left for future work.

Lastly, the degree of task differentiation in a settlement will directly correlate to the extent that labor and food shipments generate an internal bartering economy.  Task differentiation is itself largely driven by the size of the settlement, which is also critical to other parts of the economic model. Developing a Martian economy is also left for future work.

\subsection{External Martian Economy}
To examine how a Martian economy might function outside of a colony, we looked at other harsh environments with possible mineral deposits; the oceans and Antarctica. Various countries have exploited mineral deposits in the oceans before there were laws in place to govern mining or for the protection of marine vegetation and wildlife. There is a call now to bring together the fragmented efforts and explicitly address the protection of ocean ecosystems.\cite{Mengerink}

Exploration of Antarctica has gone differently given the lack of obvious mineral deposits and the harsher environment. After the initial exploration period during colonial times, there was a scientific period when all activity on Antarctica was frozen and all signers of the Antarctic Treaty agreed to collaborative, scientific exploration and experiments were allowed.\cite{Triggs} With the possibility of mineral discovery, efforts are now underway to expand on the original treaty to address additional uses.\cite{Chang}

\subsection{Energy Sources in Space}
Energy sources are another key consideration for colonization. Solar power is limited by daylight hours, seasonal variability, and dust accumulation on solar panels.\cite{JPL} We chose to model our colony on the nuclear power generator employed by NASA and the U.S. Department of Energy for the Perseverance rover in 2011. The Perseverance uses a Multi-Mission Radioisotope Thermoelectric Generator (MMRTG) power system to provide a steady supply of electricity. The MMRTG is expected to operate for at least 14 years, which is more than seven times longer than the prime mission.\cite{JPL}

Nuclear fission can be utilized to provide surface power on the moon and Mars, and small, light fission reactors could provide up to 10 kilowatts of electrical power for at least 10 years, according to a NASA website. This would provide enough electricity to power several households. Together NASA and DOE plan to design and test one of these low-enriched uranium systems on the moon by the late 2020s.\cite{Derr} For our Mars settlement we will assume a safe, stable nuclear power generator.

\subsection{Radiation}
International space scientists have written on the threat particle radiation poses to future human colonists of Mars. After four years on Mars, colonists would be exposed to a level of radiation above what is assessed safe for humans, according to the article.\cite{Young}Additionally the mission should depart for Mars when solar activity is at its peak (the solar maximum), because the increased solar activity deflects the most dangerous particles from distant galaxies.

\section{ABM description and ODD}
\subsection{KEY ASSUMPTIONS}
Due to the complexity of the model, a variety of base assumptions were made:
1) Assume the Mars colony has already been constructed, that food, air and water are being produced locally and will not require a start-up time to create.
2) Assume a nuclear generator, similar to the Perseverance power source, has been installed and the habitat has a steady electricity source for at least 7 years.

\subsection{PROBLEM FORMULATION}
The main goal of the model is to simulate a theoretical habitat environment on Mars for a colony carrying out the mining of minerals to be sent back to Earth. There is an emphasis on the mental state of each Martian agent, since that directly impacts the success of the mission. 

\begin{figure}[htp]
    \centering
    \includegraphics[width=8cm]{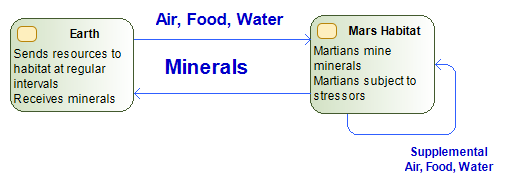}
\end{figure}

\subsection{MODEL DESCRIPTION}
\textbf{1. Purpose}

 The agent-based model MARS-COLONY is intended to provide insights into the possible scenarios that might develop from varying initial population sizes, personality type and skill levels of colonists sent to run a mining colony for minerals. The goal is to aid decision-makers with planning by highlighting potential pit falls. To make the model as realistic as possible, research is used on the studied coping capacities of common personality types, and random accidents are introduced both transporting food from Earth, as well habitat disasters on Mars that also lower food supplies. Accidents not only threaten food supplies, but considerably increase stress levels of colony members.

The model was developed using NetLogo 6.2. 

\textbf{2. Entities, State Variables, Scales}
There are two collectives and two agent types in the model. The first agent type is the Martian, which represents a single human member of the settlement. The entire group of Martian agents form the collective “Martians”. The second agent type is the stressor, which can be triggered by a settlement habitat accident or an earth shipping disaster. When triggered, the stressor interacts with the Martians in the settlement to reduce their coping capacity and health. A stressor can dissipate over time, or can be removed by the Martian agents if they succeed a skills check to recover from the habitat accident. Table 1 lists the attributes and starting attribute values for Martian agents, and Table 2 lists the attribute and starting values for stressor agents.

\begin{table}[htp]
\centering
 \begin{tabular}{|p{.33\linewidth}|p{.33\linewidth}|p{.33\linewidth}|} 
    \hline
Attribute Name & Description & Starting Value (or Range)\\
    \hline
coping-capacity & psychological coping capacity of agents & 0.84 – 0.98 \\
    \hline
resilience & coping category & “neurotic”, “reactive”, “social”, or “agreeable”\\
    \hline
skill-1	& value of skill 1	& 0 - 100\\
    \hline
skill-2	& value of skill 2	& 0 - 100\\
    \hline
health	& agent health value &	100\\
    \hline
food & food available for the given agent & 1\\
    \hline
air & air available for the given agent & 1\\
    \hline
water & water available for the given agent & 1\\
    \hline
waste & waste produced by the given agent & 1\\
    \hline
partner & the martian with which the agent is partnered, or "nobody" if not partnered & nobody\\
    \hline
taskmate-food & taskmate for food production & nobody\\
    \hline
taskmate-water & taskmate for water production & nobody\\
    \hline
taskmate-air & taskmate for air production & nobody\\
    \hline
taskmate-accident & taskmate for habitat accident recovery & nobody\\
    \hline
taskmate-waste & taskmate for waste removal & nobody\\
    \hline
\end{tabular}
\caption{Agent Attributes: Martian}
\end{table}

\begin{table}[htp]
\centering
 \begin{tabular}{|p{.33\linewidth}|p{.33\linewidth}|p{.33\linewidth}|} 
    \hline
    Attribute Name & Description & Starting Value (or Range)\\
    \hline
    identity & two types of stressor identity externalities & “Shipping” or “Habitat” \\
    \hline
\end{tabular}
\caption{Entity Attributes: Stressor}
\end{table}

\begin{figure}[htp]
    \centering
    \includegraphics[width=4cm]{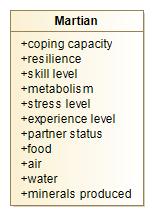}
    \caption{Martian Attributes}
\end{figure}

The settlement acts as additional entity, capable of storing air, food, water, and waste, and having a set production capacity for these resources. Martian agents interact with the settlement by working together to produce resources, store resources, and consume settlement resources. The settlement is a 50x50 grid, with each grid cell having an individual resource production capacity, but collective and unlimited storage. The geospatial location of each Martian agent is randomly assigned during model initialization, and agents move randomly around the grid. Table 3 lists the attributes and starting values for the Martian settlement. 

\begin{table}[htp]
\centering
 \begin{tabular}{|p{0.33\linewidth}|p{0.33\linewidth}|p{0.33\linewidth}|} 
    \hline
Attribute Name & Description & Starting Value (or Range)\\
    \hline
    p-food & the rate at which food, on average, can be grown per time step in each patch of the settlement & 0.5 \\
    \hline
    p-air & air recycled per patch of the settlement & 5.88\\
    \hline
    p-water & water recycled per patch of the settlement & 28\\
    \hline
    p-waste & non-recyclable waste (i.e., fecal matter, plant emissions) processed per patch of the settlement & 0\\
    \hline
    settlement-air & total air available for settlement & 5.88 * number-of-martians * 156\\
    \hline
    settlement-water & total water available for settlement & 28 * number-of-martians * 156\\
    \hline
    settlement-food & total food available for settlement & 10.5 * number-of-martians * 156\\
    \hline
    settlement-waste & total waste produced by settlement & 0\\
    \hline
    settlement-minerals & total minerals available for settlement & 0\\
    \hline
\end{tabular}
\caption{Entity Attributes: Settlement}
\end{table}

The environment is the final entity, which determines the probability of a habitat accident, earth shipment frequency and success probability, time scale, technological efficiency, and production skill requirements. Table 4 lists the environmental attributes and starting values.

\begin{table}[htp]
\centering
\begin{tabular}{|p{0.33\linewidth} | p{0.33\linewidth} |p{0.33\linewidth}|}
    \hline
    Attribute Name & Description & Starting Value (or Range)\\
    \hline
    ticks & time step for model & 1 tick = 1 week\\
    \hline
    minerals-shipment & amount of minerals received in each shipment from Earth & 100\\
    \hline
    food-shipment & amount of food received in each shipment from Earth & 10.5 * number-of-martians * 78\\
    \hline
    shipment-frequency & frequency of shipments from Earth & 78 ticks\\
    \hline
    technology & technological efficiency & 0.5\\
    \hline
    energy & energy (currently unused) & 1.0\\
    \hline
    food-s1 & amount of skill 1 required to produce food & 0 - 100\\
    \hline
    food-s2 & amount of skill 2 required to produce food & 0 - 100\\
    \hline
    water-s1 & amount of skill 1 required to produce water & 0 - 100\\
    \hline
    water-s2 & amount of skill 2 required to produce water & 0 - 100\\
    \hline
    air-s1 & amount of skill 1 required to produce air & 0 - 100\\
    \hline
    air-s2 & amount of skill 2 required to produce air & 0 - 100\\
    \hline
    waste-s1 & amount of skill 1 required to produce waste & 0 - 100\\
    \hline
    waste-s2 & amount of skill 2 required to produce waste & 0 - 100\\
    \hline
    accident-s1 & amount of skill 1 required to recover from a habitat accident & 0 - 100\\
    \hline
    accident-s2 & amount of skill 2 required to recover from a habitat accident & 0 - 100\\
    \hline
\end{tabular}
\caption{Entity Attributes: Environment}
\end{table}

\textbf{3. Process overview and scheduling}
The only user input parameter currently in the model is the initial number of Martians, which ranges from 4 – 152, in increments of 4. When the model is initialized, the total number of Martians are divided equally between four possible psychological categories: neurotic, reactive, social, and agreeable. Martians, Stressors, the Settlement, and the Environment are initialized with the values listed in Tables 1 – 4. 

At each tick, Martians are able to move, sleep, team up with other Martians, produce resources, consume resources, and engage socially. To move, the Martians face a random direction and move forward one grid cell. Next, they receive additional health from sleeping, with total health capped at 100. Martians are then able to partner with other Martians to perform a skills check for production. If the skills check is successful, they produce food, water, and air, each with a unique skills check. Waste is produced as a by-product of successful production. After resources are produced, they are then consumed. Each Martian has a weekly requirement for food, water, and air. If an individual Martian produces enough resources to meet this requirement, they consume the defined amount and any remainder is contributed to the settlement resource pool. If the Martian did not produce enough resources to meet the weekly requirement, they consume the required resource amount from the settlement resource pool. If the settlement resource pool does not have enough resources to meet a Martian’s weekly requirement, the Martian’s health level is reduced. Once the Martian’s health reaches zero, the Martian dies and is removed from the model. Additionally, there is a small random chance of death for each Martian to represent unforeseen mortality. 

The settlement restores its production capacity for each grid cell and may consume resources at each tick. Martian production is limited by the total capacity of each settlement grid cell, and reduces the amount of producible resources on that cell during the time step. At the next time step, the resources are renewed to maintain a set weekly production capacity. The settlement may also consume minerals, which are received from earth shipment, to improve technological efficiency, which then improves Martian production capability. 

At each time step, the environment has a chance to add new Martians, experience an earth shipping disaster, receive an earth shipment,  and to experience a habitat accident. Every 78 weeks, there is a chance that 4 new Martians may be added to the colony from thr resupply shuttle, with a random psychological category assigned upon addition. Also at 78 weeks, an earth shipment may be received or there is a small random chance that an earth shipping disaster may occur. If the earth shipment is received, the settlement receives food and minerals, and improves technological efficiency. If an earth shipping disaster occurs, the shipment is not received, and an earth shipping disaster stressor is generated. There is also a small random chance of a habitat accident occurring at each time step. If a habitat accident does occur, it randomly affects either the food, water, air, or mineral settlement resource supplies, and reduces the selected resource by half. It also generates a habitat accident stressor. 

If a stressor is generated, it has a chance to interact with Martians or to dissipate over time. Habitat accident stressors persist until they dissipate or are removed by a successful Martian skill check. While active, they continue to reduce the settlement resource supplies. After 4 time steps, the habitat accident stressor will dissipate and no longer affect the settlement resource supplies. 

\textbf{4. Design Concept}

\emph{Basic Principles}
The model was developed to explore the psychological, social, technological, economic, and logistical factors that would influence the long-term viability of a human Martian settlement. Martian settlers interact with each other while producing and consuming resources, which affects their overall health. They are divided into four personality types that determine the degree and direction of this effect. Each Martian also possesses a skillset that determines their success in producing resources for themselves and the settlement. The settlement is supplied by regular shipments from earth, which are susceptible to shipping disasters, and the settlement can experience a habitat accident, both of which negatively influence settlement supplies and overall Martian health.

\emph{Emergence}
The primary observed emergent phenomenon occurs in the decline of the Martian population. While the members of the settlement have an equal probability of being affected by lack of settlement resources, habitat accidents, or earth shipping disasters, Martians with the “neurotic” psychology die at a much higher rate than those of other psychologies, and once their population reaches a low enough level, the settlement population stabilizes. Martians with the neurotic psychology and a high coping capacity benefit the least from interaction with other Martians, and are penalized the most if they have a low coping capacity. Our results suggest that this effect is a driver of the Martian population decline, and once minimized or removed, can produce a stable settlement.

\emph{Adaptation}
The Martian agents do not have explicit adaptation, but do have implicit adaptation in the use of their assigned skills. During each time step, the model seeks to partner two Martian agents whose skills sum to the required production threshold for resources or for habitat accident recovery in the case of a habitat accident. This represents the idea that Martian settlers would work together based on their actual skills to accomplish various tasks needed to provide for the settlement, and would adapt as needed for these tasks.

\emph{Objectives}
The overall objective of the model is to produce a long-term stable Martian population. Each Martian’s objective is to partner with other Martians to produce resources and recover from habitat accidents. If Martians are unable to produce sufficient resources, the population will begin to decline, making future production more difficult due to a decline in the available skills required for production.

\emph{Sensing}
Martian agents move around the grid space and are able to sense other agents within 3 grid spaces of themselves. If there are one or more Martian agents within this range, a random agent is selected as a partner, and the agents interact to increase or decrease overall health and coping capacity, depending on current coping capacity and each agent’s psychological category.

\emph{Interaction}
In addition to interacting with agents in close spatial proximity, as described in the section above, Martian agents also interact to produce resources, recover from habitat accidents, and remove waste from the settlement. Pairs of agents are selected by the model based on their skill levels to achieve a set productivity threshold, which results in successful achievement of these activities. If a pair cannot be made to reach the threshold, the unpaired agents are not successful in production, recovery, or waste removal. Stressor agents interact with the settlement to reduce resource supplies or prevent an earth shipment from resupplying food and minerals.

\emph{Stochasticity}
Stochasticity appears in the assignment of agent skill levels and coping capacity, Martian agent movement, Martian death, Martian agent addition, earth shipping disaster, habitat accidents, and Martian agent teaming for production and interaction. When Martian agents are generated, levels for their skills are assigned from a random uniform distribution for the first skill, and the second skill then sums with the first to equal one hundred. Martian agents move randomly around the grid space, and face a small random probability of death at each time step. There are also random probabilities of Martian agents being added to the model, the occurrence of an earth shipping disaster when a shipment is due to arrive, or a habitat accident occurring, with the type of resource affected by the habitat accident also randomly selected. When agents pair up for resource production, habitat accident recovery, or waste removal, the pairing is random given the pair reaches the production threshold requirement. If agents are within three grid cells of each other, they can randomly be assigned to interact to affect health and coping capacity.

\emph{Collectives}
Martian agents form a settlement collective, drawing on a pool of shared resources and producing an overall amount of settlement waste. If an individual agent is unable to produce enough resources for their own needs in a given time step, they may draw on the collective settlement resources. If they produce excess resources, they are contributed to the settlement resource pool.

\emph{Observation}
Outputs collected from the model for observation include coping capacity statistics, number of Martian agents, total settlement resources, shipments received, shipping disasters and habitat accidents experienced, population coping and resilience, overall Martian health, and time elapsed.

\textbf{5. Initialization}

The setup procedure imports a Martian landscape of geographic files. The global variables are then checked for total water, food and air for the habitat. Also total energy and technology levels needed to run the habitat and mine minerals. Additional parameters are assigned for the frequency of shipment and habitat accidents. Every 78 ticks there is also a slight chance of a birth or death in the community. The individual Martian attributes are then calculated by dividing resources equally, and randomly assigning personality types and skill.

\textbf{6. Input Data}
Geographic Information System (GIS) files of the Mars landscape have been loaded for awareness, but no input data is used in this model. Mars exploration is still in the early stages and applicable data is not yet available.

\subsection{SETTING PARAMETERS}
One of the challenges of agent-based modeling is determining how to set the initial parameters for the model.  Empirical data are often a source of initial parameters, even in far-flung scenarios such as past civilizations or extra-planetary settlements.  Our model draws upon a number of sources to set our parameters.

Global Variables. The Earth shipment and shipment frequency variables are discussed above. This discussion also excludes shipping and accident counter variables, which are discussed under results. The other global variables include the settlement's total resource stockpiles (settlement-air, settlement-food, and settlement-water).  The values of the parameters are primarily derived from a review of the resource storage and production capabilities of the International Space Station (ISS).  NASA states that a single ISS resident consumes 5.88kg of air, 28L of water, and 10.5kg of food in a given week.\cite {NatGeo}\cite{NASA}\cite{Starr}  Our initial settlement-food, settlement-air, and settlement-water assume a ramp-up, pre-settlement series of deliveries that stockpile supplies equal to two earth shipment cycles (156 weeks/time steps).

However, there are proven technologies from the ISS to allow for the sustainable production of air and water.  Conversely, there is as yet no method for the sustainable production of food in low gravity.  As a result, we assume that earth shipments will deliver food but not water or air.  Earth shipments deliver sufficient food for one additional earth shipment cycle (10.5kg of food per settler per week). 

Our global variables also define several other rates necessary to the function of the settlement.  The technology parameter sets a rate of aggregate technological efficiency.  This term serves as a modifier to the amount of production generated by labor in the settlement.  Technology is initially set at 0.5

For this early version of the model, energy serves as a similar term that specifies the amount of resource consumption required for the settlement to function. This placeholder value is set at 1.0.

Lastly, the global variables set the skill levels necessary for settlement production functions.  An assumed score of 100 is needed across two sets of skills (-s1 and -s2) to successfully accomplish each of four tasks (food, water, air production, and accident recovery). Each pair of scores related to resource production is set with the same values.  For example, food-s1 is set as a random integer between 0 and 100.  Food-s2 is set as the difference between 100 and food-s1 (100 - food-s1).  Accident recovery sets both both accident-s1 and accident-s2 as a random integer from 0 to 100. This represents the ex ante uncertainty about what types of skills may be needed for a given emergency.  It is important to remember that these values set the required skill score that must be met for successful production; they do not guarantee that settlers will have these skills. 

Settler Variables. Settlers are assigned two skills.  Skill 1 is set as a random number from 0 to 100.  Skill 2 is set as 100 - Skill 1, such that each settler has a total skill level of 100.  Settlers have a partner variable (all settlers begin unpartnered) and indicator variables that store their task assignment.  

Each settler is created with one of four resilience types: nuerotic, reactive, social, and agreeable.  Each of these resilience types are assigned a coping score based on our research in team function and group theory.  These coping scores are 0.84, 0.89, 0.94, and 0.98.  Coping scores determine how settlers continue to produce and consume resources after adverse events such as a habitat accident. 

Settlement (Patch) Variables. Patch variables in this model represent the modest estimated amount of food production that could be generated in a garden or farm (0.5 kg per patch per time step), the WRPS + WTS air production system (5.88kg per patch per time step), and OPS water recycling capacity (28L per patch per time step).  These production levels are activated in the production functions only with the requisite levels of settler skill levels, as discussed above.

Exogenous Variables.  This model contains only two exogenous variables: the starting number of settlers and the number of time steps in a given model run.

\subsection{IMPLEMENTATION VERIFICATION}
 Our code shows cyclical variability in the population sizes and is able to stabilize. The code appears to run correctly, the colonists have plenty of resources, and we are able to focus on the interaction between the different personality types. However we are unable to validate the values since there has not been an actual Mars colony we can compare data to so model data serves largely to highlight problems for consideration.
 
\section{Scenarios for a Stable Mars Settlement}
\subsection{A Stable Population Size}
Our goal was to see which initial population sizes lead to a stable colony size. We did 5 runs of our model, for 28 Earth years, varying the initial population size from 10 to 170 by steps of 10. Given that there are 4 critical tasks that are needed continuously (air, water, food production and waste removal) in addition to handling disasters, and two skills needed for each task, we chose a population size of 10 as the minimum needed for a "stable" colony size. The population is allowed to dip below 10 as long as it bounces back within 1.5 years, or the amount of time between Earth resupply shuttles.  A plot of our runs below shows that all initial population sizes above 50 were able to maintain a population above 10 for all time steps. We will therefore only consider initial populations below 50 as we try to determine a minimum initial population which is able to maintain, or bounce back quickly to, a stable colony size equal to or greater than 10 for all 28 years.

\begin{figure}[htp]
    \centering
    \includegraphics[width=8cm]{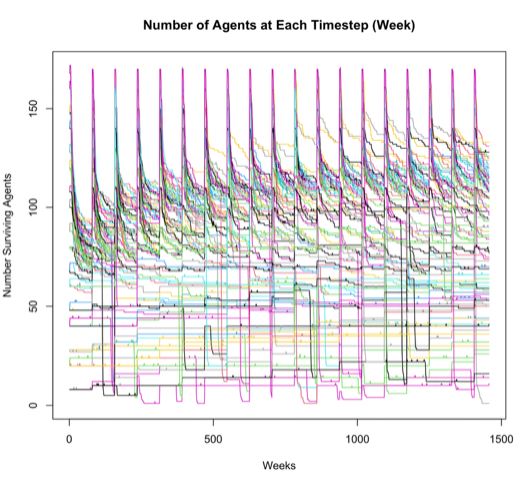}
    \caption{Variation in number of agents at each time Step (one week) in the simulation.}
\end{figure}

Since our 4 personality types are equally distributed at setup, for our focus on smaller initial populations we started at 10 (the minimum for a stable colony), then increased to 50 with a step size of 4. Each time tick in our model represents a week, so 1.5 years is equivalent to 84 ticks.

For an initial population of 10, the plot below shows the population sizes over 28 years with prolonged time periods when the population drops below 10

\begin{figure}[htp]
    \centering
    \includegraphics[width=8cm]{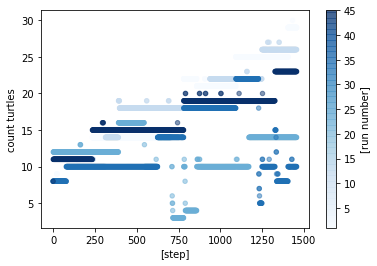}
    \caption{Population sustainability with a start of population = 10.}
\end{figure}

For an initial population size of 22, the plot below show the population sizes over 28 years remains above 10, meeting our definition of a stable colony. 

\begin{figure}[htp]
    \centering
    \includegraphics[width=8cm]{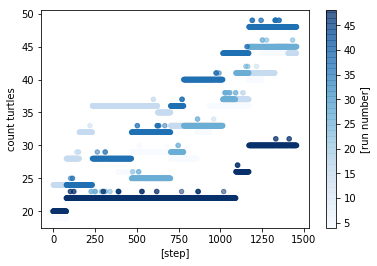}
      \caption{Population sustainability with a start of population = 22.}
\end{figure}

A summary of which small initial population sizes resulted in stable colonies over 28 years, in 5 runs, is given in the table below. For the purpose of this model, an initial population of 22 was the minimum required to maintain a viable colony size over the long run.

\begin{table}[htp]
\centering
\begin{tabular}{|p{0.33\linewidth} | p{0.33\linewidth}|p{0.33\linewidth}|}
    \hline
    Initial Population Size & Does Population Size Bounce Back Above 10 or Not\\
    \hline
    10 & No Bounce Back \\
    \hline
    14 & No Bounce Back\\
    \hline
    18 & No Bounce Back\\
    \hline
    22 & Successful Bounce Back\\
    \hline
    26 & No Bounce Back\\
    \hline
    30 & Successful Bounce Back\\
    \hline
    34 & Successful Bounce Back\\
    \hline
    38 & No Bounce Back\\
    \hline
    42 & Successful Bounce Back\\
    \hline
    46 & Successful Bounce Back\\
    \hline
    50 & Successful Bounce Back\\
    \hline
\end{tabular}
\caption{Small Initial Populations Which Do and Do Not Lead to Stable Colonies}
\end{table}

\subsection{An Agreeable Personality Type Does Best}
In all runs, the Agreeable personality type was the only one to survive the full duration of model runs. This is likely because it has the highest coping capability, and after long periods of time every agent has been exposed to a series of stressor interactions, as well as space and habitat accidents. The picture of the MARS-COLONY NetLogo model below includes a visualization of total personality types over time, and the yellow Agreeable personality type is most resilient, the Neurotic personality type was the most likely to fail, the Reactive and Social personality types alternated in between. While this model assigns equal numbers of each personality type, future work could try adjusting the proportion of each to possibly lead to a lower required minimum initial population. For example, a crew of all Agreeable personalities may be more successful.

\begin{figure}[htp]
    \centering
    \includegraphics[width=12cm]{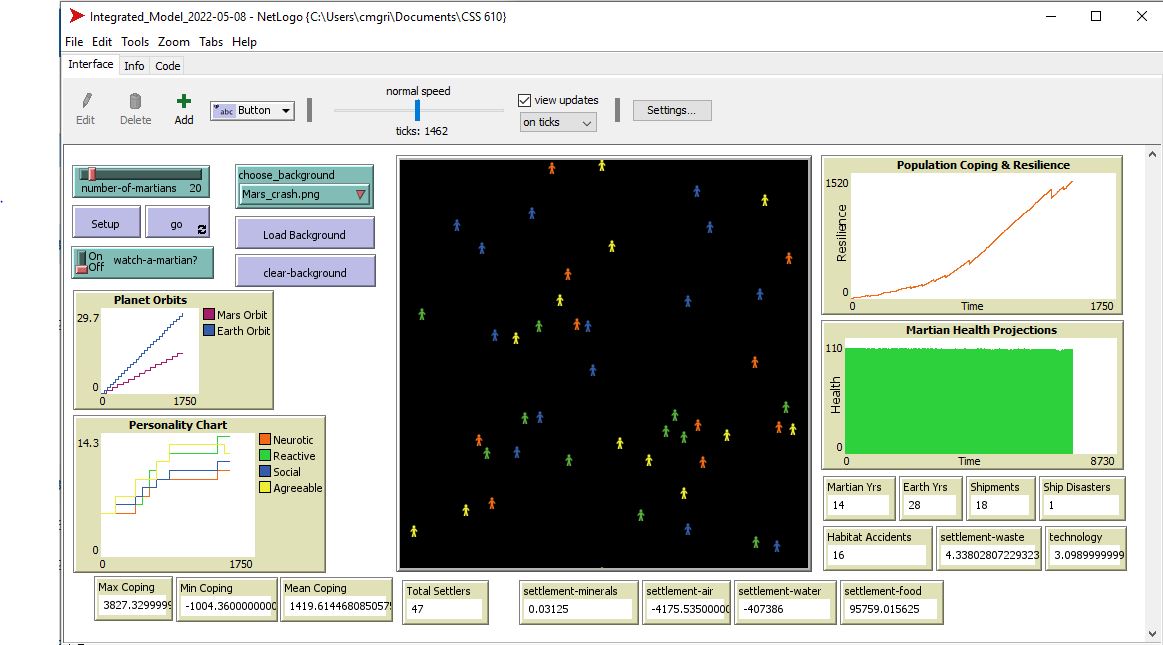}
      \caption{Screenshot of the simulation in NetLogo.}
\end{figure}

\subsection{Maintaining Sufficient Resources Despite Accidents}
Our data analysis shows that our food supplies are more than sufficiently resupplied with earth shipments, for the majority of time steps, even in scenarios where there are shipping disasters. Since the focus of this study was to examine the interplay of the 4 personality types we did not assume any production of air, water or food on the colony. Instead we provided multiple years of initial supplies, with resupplies on incoming shuttles. Similarly we were not concerned with the amount of waste generated by the colonists. Future work can focus on this area of the model and explore the trade off between resources needed from Earth, and amounts that could be generated on Mars. This is a vital component to establishing a stable colony which adds to the complexity of the model.

\section{Conclusion}
In summary, the MARS-COLONY Agent-Based Model offers a preliminary look at necessary conditions to establish a stable mining colony on Mars. The main focus is on the personality types of colonists selected and how they perform throughout their time on Mars, using their skills to mine minerals and react to random resupply shuttle accidents or habitat disasters. The stress caused by accidents, as well as from interacting with other colonists, takes a toll and Agreeable personality types were assessed to be the most enduring for the long term, whereas Neurotics showed least adaptation capacity. This study assumed a steady available nuclear energy supply and sufficient initial air, water and food with sufficient resupply in order to focus on personality and mental health of colonists, however those are recommended areas of future work.

\printbibliography

\end{document}